\documentclass[12pt]{article} 

\usepackage{epsfig} 
\usepackage{amsfonts}
\usepackage{amssymb}
\usepackage{amsmath}  
\usepackage{amsbsy}  
\usepackage{wasysym}
\newlength{\largfig}
\def\beq{\begin{equation}} 
\def\eeq{\end{equation}} 
 
\def\bea{\begin{eqnarray}} 
\def\eea{\end{eqnarray}}

\def\al{\alpha} 
\def\be{\beta}

\def\wwjj{W^+W^-jj}

\def\wpzlldec{e^+\nu_e \,\mu^+\mu^-}
\def\wmzlldec{e^-\bar\nu_e \,\mu^+\mu^-}
\def\wpzjj{W^+Zjj}
\def\wmzjj{W^-Zjj}
\def\wzjj{W^\pm Zjj}

\def\zzjj{ZZjj}

\def\mc{\mathcal}
\def\mr{\mathrm}

\def\MB{{\cal M}_B}
\def \ep{\epsilon} 
\def \eps{\epsilon} 
\def\lq{\left[} 
\def\rq{\right]}

\def\({\left(} 
\def\){\right)} 
\def\Re{\mathop{\rm Re}} 

\def\sla#1{\ifmmode% 
\setbox0=\hbox{$#1$}% 
\setbox1=\hbox to\wd0{\hss$/$\hss}\else% 
\setbox0=\hbox{#1}% 
\setbox1=\hbox to\wd0{\hss/\hss}\fi% 
#1\hskip-\wd0\box1 }

\def\non{\nonumber}

\newskip\humongous \humongous=0pt plus 1000pt minus 1000pt

\newif\ifdtup

\catcode`@=12 
 
\catcode`\@=11 
 
\@addtoreset{equation}{section} 

\def\theequation{\thesection.\arabic{equation}} 
 
\def\@normalsize{\@setsize\normalsize{15pt}\xiipt\@xiipt 
\abovedisplayskip 14pt plus3pt minus3pt% 
\belowdisplayskip \abovedisplayskip 
\abovedisplayshortskip \z@ plus3pt% 
\belowdisplayshortskip 7pt plus3.5pt minus0pt} 
 
\def\small{\@setsize\small{13.6pt}\xipt\@xipt 
\abovedisplayskip 13pt plus3pt minus3pt% 
\belowdisplayskip \abovedisplayskip 
\abovedisplayshortskip \z@ plus3pt% 
\belowdisplayshortskip 7pt plus3.5pt minus0pt 
\def\@listi{\parsep 4.5pt plus 2pt minus 1pt 
     \itemsep \parsep 
     \topsep 9pt plus 3pt minus 3pt}} 
 
\@twosidetrue

\catcode`\@=11  
\def\section{\@startsection{section}{1}{\z@}{3.5ex plus 1ex minus 
   .2ex}{2.3ex plus .2ex}{\large\bf}}

\def\thesection{\arabic{section}} 
\def\thesubsection{\arabic{section}.\arabic{subsection}} 
\def\thesubsubsection{\arabic{section}.\arabic{subsection}.\arabic{subsubsection}} 
 
\def\appendix{\setcounter{section}{0} 
 \def\thesection{\Alph{section}} 
 \def\theequation{\Alph{section}.\arabic{equation}} 
\def\thesubsection{\Alph{section}.\arabic{subsection}} 
\def\thesubsubsection{\Alph{section}.\arabic{subsection}.\arabic{subsubsection}} 
 
\def\section{\@startsection{section}{1}{\z@}{3.5ex plus 1ex minus 
   .2ex}{2.3ex plus .2ex}{\large\bf}} 
}

\newcommand{\ccaption}[2]{ 
  \begin{center} 
    \parbox{0.85\textwidth}{ 
      \caption[#1]{\small\it {#2}}} 
  \end{center}    } 
 
\def \ep{\epsilon} 
\def \eps{\epsilon} 
\def \to   {\mbox{$\rightarrow$}}

\newcommand\sigmac{$\sigma_{\rm cuts}$}
\newcommand\sigmacNLO{$\sigma_{\rm cuts}^{\rm NLO}$}
\newcommand\sigmacLO{$\sigma_{\rm cuts}^{\rm LO}$}

\newcount\minutes 
\newcount\scratch 
 
\def\timestamp{% 
\scratch=\time 
\divide\scratch by 60 
\edef\hours{\the\scratch} 
\multiply\scratch by 60 
\minutes=\time 
\advance\minutes by -\scratch 
---$\,$\hours:\null 
\ifnum\minutes< 10 0\fi 
\the\minutes}

 \newcommand\tol{\,\to\,}
 
\begin{document} 
\begin{titlepage} 
\nopagebreak 
{\flushright{ 
        \begin{minipage}{5cm} 
	 Bicocca--FT--07--1  \\
         KA--TP--01--2007  \\       
         SFB--CPP--07--02  \\       
        {\tt hep-ph/0701105}\hfill \\ 
        \end{minipage}        } 
 
} 
\vfill 
\begin{center} 
{\LARGE \bf \sc 
 \baselineskip 0.9cm 
Next-to-leading order QCD corrections to $W^+Z$ and $W^-Z$ production via\\
vector-boson fusion 

} 
\vskip 0.5cm  
{\large   
Giuseppe Bozzi$^1$, Barbara J\"ager$^{1,2}$, Carlo Oleari$^3$ 
and Dieter Zeppenfeld$^1$ 
}   
\vskip .2cm  
{$^1$ {\it Institut f\"ur Theoretische Physik, 
        Universit\"at Karlsruhe, P.O.Box 6980, 76128 Karlsruhe, Germany}
}\\ 
{$^2$ {\it RIKEN, Radiation Laboratory, 2-1 Hirosawa, Wako, 
	351-0198 Saitama, Japan}}\\   
{$^3$ {\it 
        Universit\`a di Milano-Bicocca and INFN Sezione di Milano-Bicocca
        20126 Milano, Italy}}\\   
 
 \vskip 
1.3cm     
\end{center} 
 
\nopagebreak 
%\vfill 
%\vskip 3cm 
\begin{abstract}
We present the calculation of the next-to-leading order QCD corrections to
electroweak  
$pp\tol \wpzlldec jj$ and $pp\tol \wmzlldec jj$ production at the 
CERN LHC in the form of a fully flexible parton-level Monte Carlo program. 
The QCD corrections to the total cross sections are modest, changing the
leading-order
results by less than 10\%. At the Born level, the shape of kinematic 
distributions can depend significantly on the choice of factorization scale. 
This theoretical uncertainty is
strongly reduced by the inclusion of the next-to-leading order QCD corrections.
\end{abstract} 
\vfill 
\today \timestamp \hfill 
\vfill 

% PACS: 14.80.Bn
\end{titlepage} 
\newpage               
%
%%%%%%%%%%%%%%%%%%%%%%%%%%%%%%%%%%%%%%%%%%%%%%%%%%%%%%%%%%%%%%%%%
%                                                       
\section{Introduction}
\label{sec:intro}
One of the primary goals of the CERN Large Hadron Collider (LHC) is a detailed
understanding of the mechanism of electroweak (EW) symmetry breaking. In this
context, vector-boson fusion (VBF) processes constitute an interesting class of
reactions. Higgs production in VBF (i.e., the reaction $qq\tol qqH$) is 
discussed as a possible discovery mode of a Standard Model Higgs
boson~\cite{ATLAS,CMS,VBF:H}.  
Once the Higgs boson has been found, VBF will be used to determine its 
CP properties~\cite{VBF:CP} and constrain its couplings to gauge bosons and 
fermions~\cite{VBF:C}. 
In the absence of a light Higgs boson, gauge-boson pair-production in 
VBF is of particular importance, as this reaction  
gives access to the scattering of longitudinally polarized vector bosons
via the subprocess $V_L V_L\tol V_L V_L$ and is therefore intimately related 
to the mechanism of electroweak symmetry breaking (here
$V$ refers to a $W^\pm$ or a $Z$ boson). Typical signatures of ``new
physics'', such as strong electroweak symmetry breaking, show up 
as enhancements over the predicted Standard Model production rates
in the cross sections for EW $qq\tol qqVV$ production at 
high center-of-mass energies.
Accurate predictions for EW $VV\,jj$ production are therefore needed not
only in estimates of backgrounds to the Higgs signal in VBF, 
but also for the search of
physics beyond the Standard Model~\cite{SEWSB}. 

In a set of previous publications, we have presented next-to-leading order (NLO)
QCD predictions for 
$\wwjj$ and $\zzjj$ production in VBF~\cite{JOZ:WW,JOZ:VV} in the form of flexible
parton-level Monte Carlo programs, allowing for the calculation of kinematic
distributions and the implementation of realistic
experimental acceptance cuts. With the present article, we extend 
our work to include EW $\wzjj$ production. 

$\wzjj$ production via VBF at hadron colliders was considered 
in~\cite{DHT:1991} in the framework of the
``effective $W$ approximation'' (EWA)\cite{EWA}. 
The EWA treats the incoming gauge bosons
as on-shell particles and does not give access to the kinematic
distributions of the final-state jets characteristic for VBF-type reactions.
Going beyond the EWA, 
a full tree-level calculation of $qq\tol qqWZ$ processes, including $W$ and $Z$
leptonic decay correlations, has been presented in~\cite{Bar:1992} and in turn
been applied to the investigation of strongly-interacting 
gauge-boson systems in Ref.~\cite{Bag:WW}. 

We extend these calculations by presenting the complete NLO QCD
predictions for $pp\tol \wpzlldec jj$ and $pp\tol \wmzlldec jj$. We consider all
resonant and non-resonant Feynman diagrams giving rise to the specified leptonic
final state by $t$-channel weak-boson exchange at order 
$\mc{O}(\alpha_s\alpha^6)$. Contributions from weak-boson exchange in the
$s$-channel are strongly suppressed in the phase-space regions where VBF can be
observed experimentally and therefore disregarded throughout. Gauge invariance
requires us not only to consider ``true'' VBF diagrams, where the leptons are
produced via gauge-boson exchange in the $t$-channel, but also graphs with
one or two gauge bosons emitted from either quark line 
(see Fig.~\ref{fig:feynBorn}). We do not specifically require the charged leptons to
stem from a $Z$ boson, but also take $\gamma\tol\ell^+\ell^-$ contributions
into 
account. For simplicity, we refer to the $pp\tol \wpzlldec jj$ and  
$pp\tol \wmzlldec jj$ processes computed in this way as ``EW $\wzjj$''
production in the following.

The outline of this article is as follows.  
The technical framework of our calculation is described in Sec.~\ref{sec:calc}. 
In Sec.~\ref{sec:res} we
discuss the phenomenological results obtained with the parton-level Monte Carlo
program which we have developed. Conclusions are given in Sec.~\ref{sec:concl}.

%
%%%%%%%%%%%%%%%%%%%%%%%%%%%%%%%%%%%%%%%%%%%%%%%%%%%%%%%%%%%%%%%%%
%                                                       
%
\section{Calculational framework}
\label{sec:calc}
The computation of NLO QCD corrections to EW $\wzjj$ production is 
performed in complete
analogy to our earlier work on $\wwjj$ and $\zzjj$ production in VBF
\cite{JOZ:WW,JOZ:VV}. 
For clarity, we briefly repeat the basic steps of the calculation and focus on
the aspects specific to $\wzjj$ production in the following.  

\begin{figure}[!htb] 
\centerline{ 
\epsfig{figure=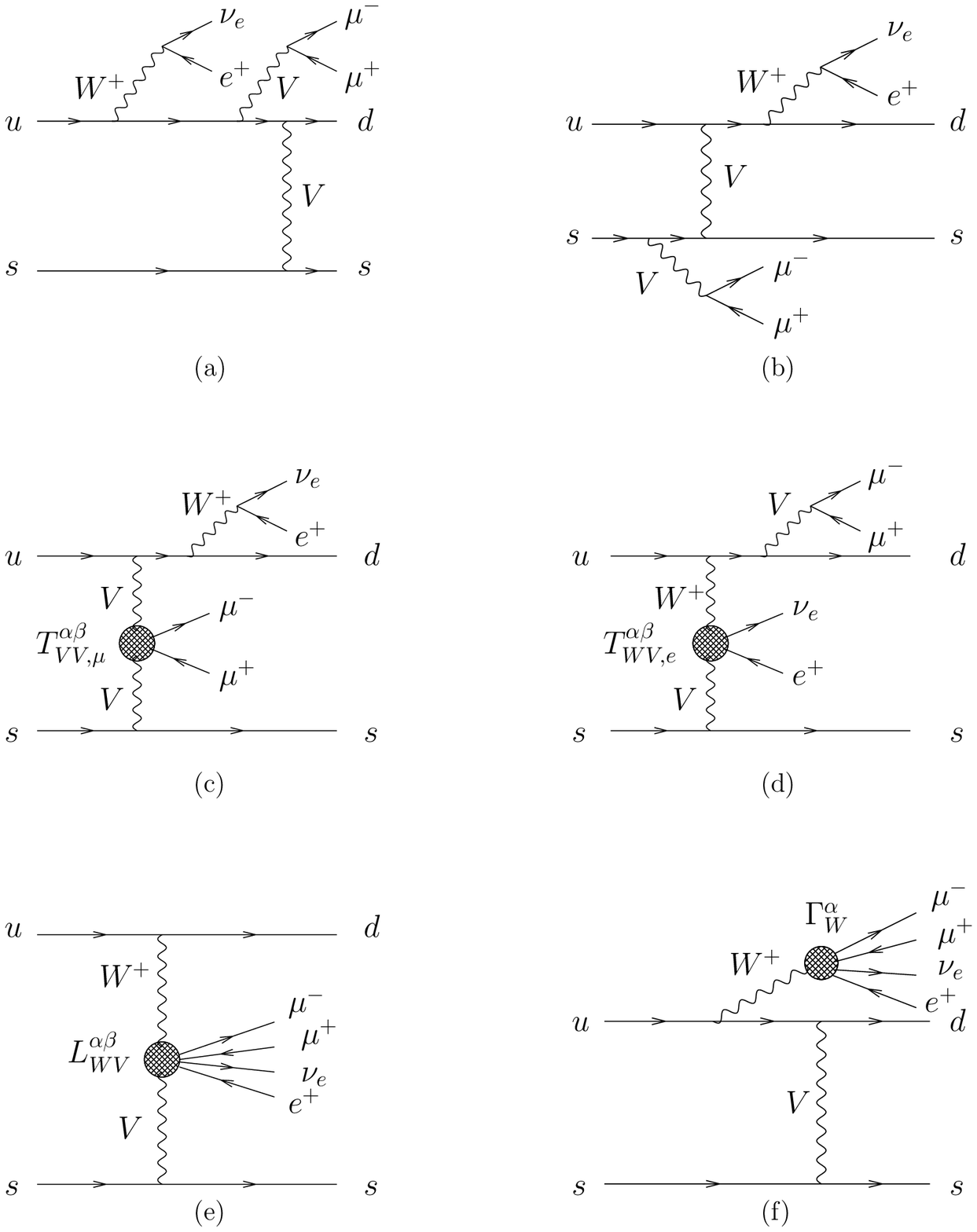,width=0.93\textwidth,clip=}
} 
\ccaption{} 
{\label{fig:feynBorn} 
The six Feynman-graph topologies contributing to the Born process $us\,\to\,
ds\,\wpzlldec$. Not shown are the diagrams analogous
to~(a), (b), (c), (d), and~(f), with $W^+$ and/or $V$ emission off the lower
quark 
line and $t$-channel $W$ exchange (this one is illustrated in
Fig.~\protect{\ref{fig:t-channel-W}}).   
$V$ denotes a $Z$ boson or a photon.
}
\end{figure}

For the evaluation of partonic matrix elements  we employ the
amplitude techniques of~\cite{HZ}, supplemented by the use of ``leptonic
tensors'' as elaborated in some detail in Refs.~\cite{JOZ:WW,JOZ:VV}. 
In the case of $pp\to\wpzlldec jj$ scattering, 210 diagrams contribute to any 
subprocess with a specific combination of quark flavors at leading
order~(LO). A few  
representative graphs for the channel $us\tol ds\,\wpzlldec$ are depicted in
Fig.~\ref{fig:feynBorn}. Here and in the following, $V$ labels $\gamma$ or 
$Z$ exchange.
All diagrams can be grouped in six different topologies, which correspond to
the following configurations: 
\begin{itemize}
\item  two external vector bosons are emitted from the same  
quark line and subsequently decay into lepton
pairs [topology~(a)].  
\item  two external vector bosons are emitted from different quark 
lines and subsequently decay into lepton
pairs [topology~(b)]. This topology also includes $t$-channel $W$ exchange
and $W^+$ emission off the $s$ quark line. 
\item one external vector boson is emitted from a quark line, while the
production of the remaining lepton pair is described by the sub-amplitudes 
$VV\to\mu^+\mu^-$, $W^+W^-\to\mu^+\mu^-$, or $W^+V\tol e^+\nu_e$ 
[topologies~(c) or~(d)], which 
are characterized by the leptonic tensors $T_{VV,\mu}^{\alpha\beta}$,  
$T_{WW,\mu}^{\alpha\beta}$, and $T_{WV,e}^{\alpha\beta}$, where 
$\al,\be$ denote the Lorentz indices carried by the internal vector bosons
and $V$ is a $Z$ boson or a photon.
\item all leptons are produced in the subprocess $W^+V\tol \wpzlldec$, 
described by the leptonic tensor $L_{WV}^{\al\be}$ [topology~(e), which is the 
vector boson fusion topology]. 
\item all leptons are produced in the subprocess 
$W^+\tol \wpzlldec$, giving rise to the tensor $\Gamma_W^\al$ [topology~(f)].
\end{itemize}
The leptonic tensors are indicated by the cross-hatched blobs in 
Fig.~\ref{fig:feynBorn} and they correspond to a sum of Feynman 
diagrams involving electroweak bosons and leptons only. 
Their use for diagrams of the same topology but with differences 
in quark propagators allows us to
organize the calculation in a modular and efficient way. Building blocks that
agree in several diagrams are computed  only once per phase-space point rather
than separately for each graph. Furthermore, a future implementation of
new-physics 
effects in the bosonic/leptonic sector is straightforward in this framework.
The partonic matrix elements for $pp\to\wmzlldec jj$  are evaluated 
in complete analogy to the $pp\tol \wpzlldec jj$ channel. 
They give rise to new leptonic tensors for the 
sub-amplitudes $W^-V\tol e^-\bar\nu_e$, $W^-V\tol \wmzlldec$ and 
$W^-\tol \wmzlldec$.

In either case, contributions from anti-quark initiated processes such as  
$u\bar s\tol d \bar s\,\wpzlldec$, obtained by crossing the above diagrams, are
fully taken into account. However, graphs comprising vector-boson exchange in
the $s$-channel with subsequent decay into a pair of jets are neglected
throughout, as they are strongly suppressed in the phase-space
regions where VBF can be observed experimentally. For identical-flavor
combinations, besides $t$-channel also $u$-channel exchange
diagrams arise. After the application of typical VBF cuts, the
interference of $t$- and $u$-channel contributions becomes negligible 
\cite{Co} and is therefore disregarded in our calculation.

The calculation of NLO QCD corrections to EW~$\wpzjj$ and $\wmzjj$ production
via VBF is based on the dipole subtraction formalism, in the version proposed
by Catani and Seymour~\cite{CS}.  Since the QCD structure of these reactions
is identical to the related case of Higgs production in VBF, the
counter-terms are of the same form and can be readily adapted from
Ref.~\cite{FOZ:Hjj}.
The real-emission contributions are computed in 
analogy to the LO matrix elements discussed above. In addition to quark and
anti-quark initiated processes with extra gluon emission, such as $us\to
dsg\,\wpzlldec$, channels with a gluon in the initial state (e.g.\ $ug\tol
ds\bar s\,\wpzlldec$) have to be considered.  
The virtual contributions,
obtained from the interference of one-loop diagrams with the Born amplitude,
are computed in the dimensional reduction scheme~\cite{DR_citation} in $d=4-2\eps$ space-time 
dimensions. They include self-energy, triangle, box and pentagon 
corrections on either the upper or
the lower quark line. Contributions from graphs  with gluons attached to 
both the upper and lower quark lines vanish  at order $\alpha_s$, within our
approximations, due to color conservation.
The interference of  all the one-loop corrections along a quark line, ${\cal
  M}_V$,  with the Born amplitude, $\MB$, yields the contribution
\bea
\label{eq:virtual_born}
2 \Re \lq {\cal M}_V\MB^* \rq
&=& |\MB|^2 \frac{\alpha_s(\mu_R)}{2\pi} C_F
\(\frac{4\pi\mu_R^2}{Q^2}\)^\epsilon \Gamma(1+\epsilon)\\ \non
 &&\times
\lq-\frac{2}{\epsilon^2}-\frac{3}{\epsilon}+c_{\rm virt}\rq\
+2 \Re \lq \widetilde{\cal M}_V\MB^* \rq \,,
\eea
where  $Q$ is the momentum transfer between the initial- and the final-state
quark,  
$\mu_R$ is the renormalization scale, $C_F=4/3$, 
$c_{\rm virt}=\pi^2/3-7$, and $\widetilde{\cal M}_V$ is a completely 
finite remainder. 
Since we have two distinct classes of virtual corrections, along the upper and
the lower quark line, we have two replicas of Eq.~(\ref{eq:virtual_born}). 
Each comes with its own distinct momentum transfer, $Q^2$.

The poles in (\ref{eq:virtual_born}) are canceled by respective singularities
in the phase-space integrated counter-terms, which in the notation of
Ref.~\cite{CS} are given by 
\beq
\label{eq:I}
\langle \mc{I}(\ep)\rangle  = |\MB|^2 \frac{\alpha_s(\mu_R)}{2\pi} C_F
\left(\frac{4\pi\mu_R^2}{Q^2}\right)^\epsilon \Gamma(1+\epsilon)
\lq\frac{2}{\epsilon^2}+\frac{3}{\epsilon}+9-\frac{4}{3}\pi^2\rq\;.
\eeq
Special care has to be taken in the numerical evaluation of
the remainder, $\widetilde{\cal M}_V$, which can be computed in
$d=4$ dimensions. We have calculated the two-, three-, and four-point tensor
integrals contributing to $\widetilde{\cal M}_V$ by a 
Passarino-Veltman reduction procedure~\cite{PV}, which is stable in the phase
space regions covered by VBF-type reactions. For pentagon contributions,
however, this technique  gives rise to numerical instabilities, if kinematical
invariants, such as the Gram determinant, become small. We have therefore
resorted to the reduction scheme proposed by Denner and Dittmaier for the
tensor reduction of pentagon integrals~\cite{DD}. 
In order to tag remaining numerical instabilities of the pentagon 
contributions, we check Ward identities between pentagon and box graphs 
for each pentagon contribution. The fraction of events for which numerical 
instabilities lead to violations exceeding 10\% can be brought down to the 
permille-level.
The respective phase-space points are disregarded for
the calculation of the finite parts of the pentagon contributions 
and the remaining
pentagon part is compensated for this loss. In Ref.~\cite{JOZ:WW}
it has been shown that this gauge check procedure ensures stable 
results for the virtual
corrections, while the error induced by the approximation is far below the
numerical accuracy of our Monte Carlo program.

In addition to the stability tests for the pentagon diagrams, 
we have performed extensive checks for the LO and the real-emission
amplitudes as well as for the total cross sections obtained with our parton-level Monte
Carlo program. We have compared the Born amplitude and the real-emission diagrams
with the fully automatically generated results provided by {\tt
  MadGraph}~\cite{madgraph} 
and found complete agreement. The real-emission contributions are QCD gauge
invariant. Finally, our total LO cross sections for EW $\wpzlldec jj$ and 
$\wmzlldec jj$ production with a minimal set of cuts agree with the respective
results obtained by {\tt MadEvent}.

%
%%%%%%%%%%%%%%%%%%%%%%%%%%%%%%%%%%%%%%%%%%%%%%%%%%%%%%%%%%%%%%%%%
%                                                       
\section{Predictions for the LHC}
\label{sec:res}
The cross-section contributions discussed above have been implemented in
fully-flexible parton-level Monte Carlo programs for EW $\wpzjj$ and $\wmzjj$ 
production  at NLO QCD accuracy in the {\tt vbfnlo} framework that has been
developed for the study of EW $Hjj$, $Zjj$, $W^\pm jj$, $\wwjj$, $\zzjj$ and
slepton pair production~\cite{FOZ:Hjj,OZ:Vjj,JOZ:WW,JOZ:VV,ZK:SL}.

We use the CTEQ6M parton distribution functions with
$\alpha_s(m_{Z})=0.118$ 
at NLO, and the CTEQ6L1 set at LO~\cite{cteq6}. In our
calculation, quark masses are set to zero throughout and contributions from
external top or bottom quarks are neglected.

For the Cabibbo-Kobayashi-Maskawa matrix, $V_{CKM}$, we have used a 
diagonal form, equal to the identity matrix, which is equivalent to employing 
the exact $V_{CKM}$, when the summation over final-state quark flavors is 
performed and when quark masses are neglected. 
\begin{figure}[tb] 
\centerline{ 
\epsfig{figure=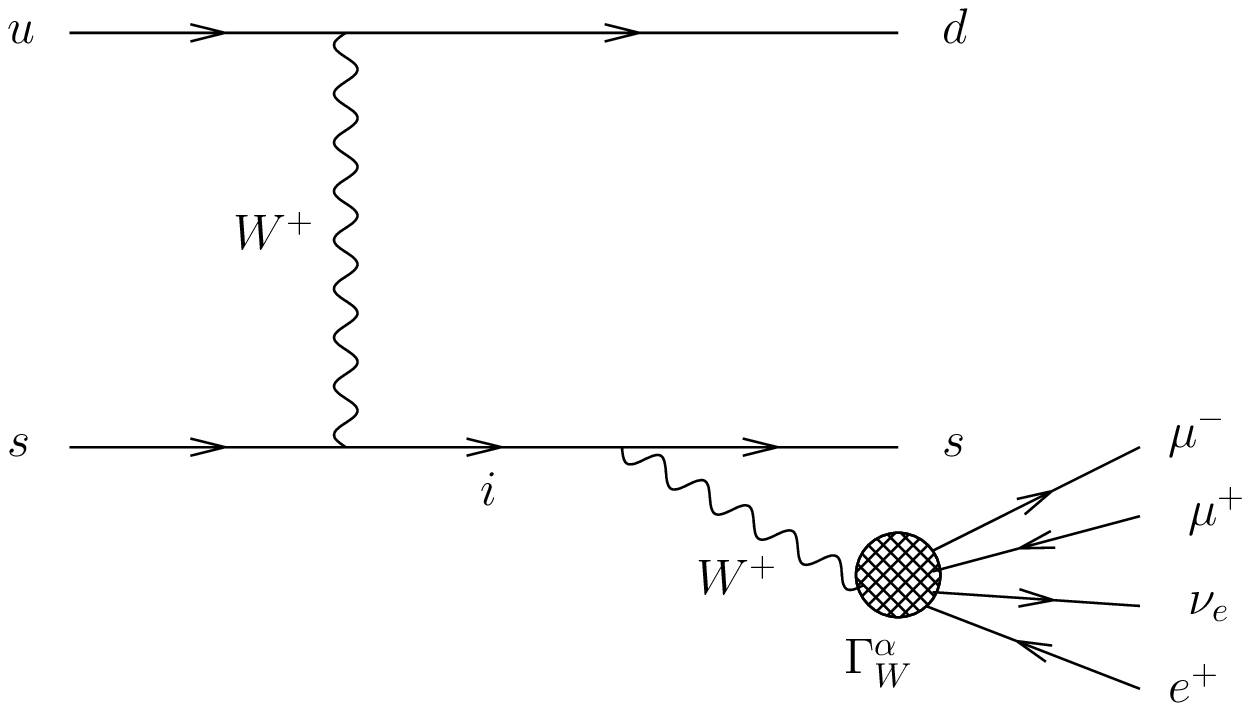,width=0.6\textwidth,clip=}
} 
\ccaption{} 
{\label{fig:t-channel-W} 
Feynman-graph topology contributing to the Born process $us\,\to\,
ds\,\wpzlldec$ via $W$ exchange in the $t$-channel. The $s$-quark on the
lower quark line is transformed into an up-type quark ($i=u,c,t$) by the
attached weak bosons.}
\end{figure} 
In fact, using the scattering $us\to ds\,\wpzlldec$ as a template, 
and referring to the topologies illustrated
in  Figs.~\ref{fig:feynBorn} and~\ref{fig:t-channel-W}, we can see that
\begin{itemize}
\item the diagrams depicted in Figs.~\ref{fig:feynBorn} are all proportional
  to $V_{ud}$
\item the diagrams depicted in Fig.~\ref{fig:t-channel-W} are proportional
  to $V_{ud} V^\dagger_{si} V_{is}$ where $i=u,c,t$. 
  In the approximation of zero
  top and charm mass, the sum $\sum_{i=u,c,t}V^\dagger_{si} V_{is}$ 
  factorizes,  and, due to unitarity, it is equal to 1. The correction 
  due to finite top quark mass is proportional to 
  $|V_{st}|^2\approx 10^{-3}$ and entirely negligible.
\end{itemize}
The squared matrix element for the scattering $us\to ds\,\wpzlldec$ is then
proportional to $|V_{ud}|^2$.  If we do not detect the final-state quark
flavors, all the processes $us\to ds\,\wpzlldec$, $us\to ss\,\wpzlldec$ and
$us\to bs\,\wpzlldec$ contribute to $pp\tol \wpzlldec jj$ scattering. The
sum over these three sub-processes is proportional to
\beq
  |V_{ud}|^2 + |V_{us}|^2 + |V_{ub}|^2
\eeq
which is equal to 1 by unitarity.
%END

In order to treat massive vector-boson propagators in a consistent and
electromagnetically gauge-invariant way, we resort to a modified version of
the complex-mass scheme~\cite{Denner:1999gp} which has already been used in
Refs.~\cite{OZ:Vjj,JOZ:WW,JOZ:VV,ZK:SL}. We globally replace gauge
boson masses in propagators by $m_V^2\tol m_V^2-im_V\Gamma_V$, while retaining
a real value for $\sin^2\theta_W$. 

We have chosen $m_Z=91.188$~GeV, $m_W=80.419$~GeV and the measured value of
$G_F=1.166\times10^{-5}/$GeV$^2$ as electroweak input parameters, from which we
obtain $\alpha_{QED}=1/132.51$ and $\sin^2\theta_W=0.2223$ using LO electroweak
relations. In order to reconstruct jets from final-state partons, we use the
$k_T$ algorithm~\cite{kToriginal,kTrunII} with resolution parameter $D=0.8$. 

We apply generic cuts that are
typical for VBF studies at the LHC~\cite{VBF:H} and require at least 
two hard jets with 
\beq
\label{eq:cutspj}
p_{Tj} \geq 20~{\rm GeV} \, , \qquad\qquad |y_j| \leq 4.5 \, ,
\eeq 
where $y_j$ is the rapidity of the (massive) jet momentum which is 
reconstructed as the four-vector sum of massless partons of pseudorapidity 
$|\eta|<5$. The two reconstructed jets of highest transverse momentum are 
called  ``tagging jets''.

In the following, we consider the specific leptonic final states
$e^+\nu_e\mu^+\mu^-$ and $e^-\bar\nu_e\mu^+\mu^-$. Results for four-lepton final
states with any relevant combination of electrons or muons and the associated
neutrino (i.e.\ $e^+\nu_e\mu^+\mu^-$, $\mu^+\nu_\mu e^+e^-$, $e^+\nu_e e^+e^-$, 
$\mu^+\nu_\mu\mu^+\mu^-$ in the $\wpzjj$ case and accordingly for $\wmzjj$
production) can, apart from negligible identical lepton interference effects,
be obtained thereof by multiplying our predictions by a factor
of four. 

To ensure that the charged leptons are well observable, we demand
\bea
&& p_{T\ell} \geq 20~{\rm GeV} \,,\qquad |\eta_{\ell}| \leq 2.5  \,,\qquad 
\triangle R_{j\ell} \geq 0.4 \, , \nonumber \\
&& m_{\ell\ell} \geq 15~{\rm GeV} \,,\qquad  
\triangle R_{\ell\ell} \geq 0.2 \, ,
\label{eq:cutspl}
\eea
where $\triangle R_{j\ell}$ and $\triangle R_{\ell\ell}$ denote the
jet-lepton and (charged) lepton-lepton separation in the rapidity-azimuthal 
angle plane and $m_{\ell\ell}$ is the invariant mass of a same-flavor 
charged lepton pair.  
In addition, the charged leptons need to fall between
the rapidities of the two tagging jets
\beq
\label{eq:cutsyl}
y_{j,min}  < \eta_\ell < y_{j,max} \, .
\eeq
Backgrounds to VBF are significantly reduced by requiring a large rapidity
separation of the two tagging jets. We therefore impose the cut
\beq
\label{eq:cutsyjj}
\Delta y_{jj}=|y_{j_1}-y_{j_2}|>4\; 
\eeq
and require the tagging jets to reside in opposite detector
hemispheres,
\beq
\label{eq:cutsh}
y_{j_1} \times y_{j_2} < 0\, ,
\eeq
with an invariant mass 
\beq
\label{eq:cutsmjj}
M_{jj} > 600~{\rm GeV}\;.
\eeq
With these cuts, the LO differential cross section for EW $\wzjj$ production via
VBF is finite, since finite scattering angles for the two quark jets are
enforced. At NLO, initial-state singularities due to collinear $q\tol qg$
and $g\tol q\bar q$ splitting can arise. These are taken care of by factorizing
them into the respective quark and gluon distribution functions of the proton.
Additional divergences stemming from the $t$-channel exchange of 
low-virtuality photons in real-emission diagrams are avoided by imposing a cut
on the virtuality of the photon, $Q_{\gamma,min}^2=4$~GeV$^2$. 
Events that do not satisfy the $Q_{\gamma,min}^2$ constraint would give rise
to a $q\rightarrow q\gamma$ collinear singularity, which is part of the
corrections to $p\gamma\to \wzjj$ and is not taken into account here.
In the related case of $W^+jj$ production via VBF~\cite{OZ:Vjj} it was found
that the NLO QCD cross section within typical VBF cuts is 
quite insensitive to the photon cutoff, changing by a mere 0.2~\% when
$Q_{\gamma,min}^2$ is lowered from 4~GeV$^2$ to 0.1~GeV$^2$. 

\begin{figure}[!tb] 
\centerline{ 
\epsfig{figure=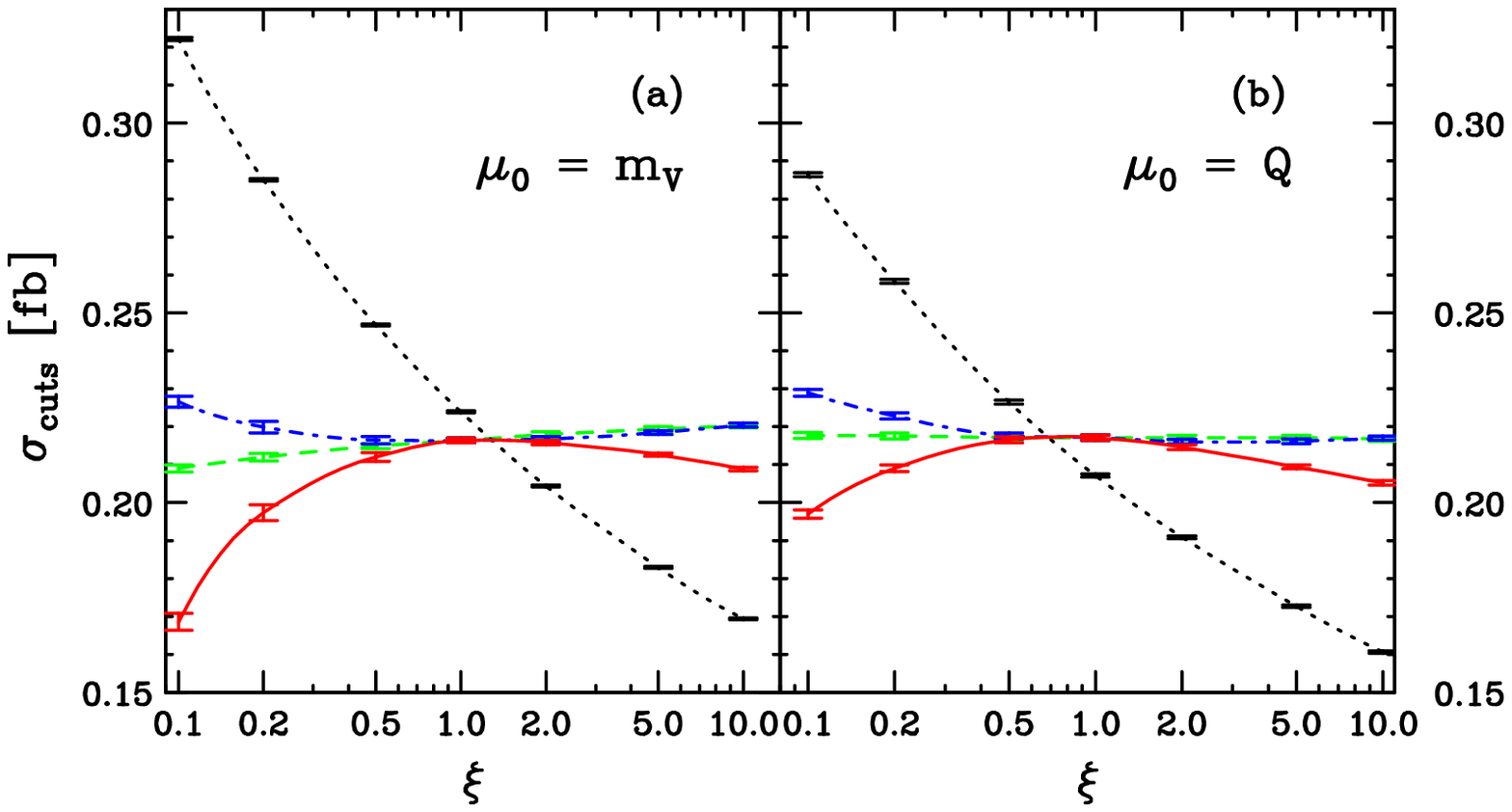,width=0.95\textwidth,clip=}
} 
\caption
{\label{fig:scale-dep} 
Dependence of the total  $pp\tol \wpzlldec jj$  cross section at
the LHC on the factorization and renormalization scales for two different
choices of $\mu_0$. 
The NLO curves show \sigmacNLO\ as a function of the scale parameter
$\xi$ for three different cases: $\mu_R=\mu_F=\xi\mu_0$ (solid red),
$\mu_F=\xi\mu_0$ and $\mu_R=\mu_0$ (dot-dashed blue), $\mu_R=\xi\mu_0$ and
$\mu_F=\mu_0$ (dashed green). The LO cross sections depend only on $\mu_F$
(dotted black). 
}
\end{figure}

The total cross section, $\sigma_{\rm cuts}$, 
for EW $\wpzjj$ production within the
cuts of Eqs.~(\ref{eq:cutspj})--(\ref{eq:cutsmjj}) and for a Higgs mass of
$m_H=120$~GeV is displayed in Fig.~\ref{fig:scale-dep} 
for different values of
the renormalization and factorization scales, $\mu_R$ and $\mu_F$, which are
taken as multiples of the mass scale $\mu_0$, 
\beq
\label{eq:scale}
\mu_R = \xi_R\,\mu_0\;,\qquad\qquad  
\mu_F = \xi_F\,\mu_0\; .
\eeq
In panel~(a), $\mu_0$ is taken as the average vector boson mass involved in
the reaction, $m_V=(m_Z+m_W)/2$. 
In panel~(b), $\mu_0=Q$, where $Q$ (see Eq.~(\ref{eq:virtual_born})) is the
momentum transfer carried by the exchanged vector boson in VBF
graphs as in Fig.~\ref{fig:feynBorn}(e), more precisely the vector boson
attached to the upper (lower) quark line for QCD 
corrections and parton densities of the upper (lower) line.
The LO cross section is of order
$\alpha_s^0\alpha^6$ and therefore depends on the factorization scale only,
which is set to $\mu_F = \xi\,\mu_0$ (dotted black curves). For \sigmacNLO\
we show three cases: $\xi_F=\xi_R=\xi$ (solid red lines), $\xi_F=\xi,
\xi_R=1$ (dot-dashed blue lines) and $\xi_F=1, \xi_R=\xi$ (dashed green
lines).
In the range  $0.5\le \xi\le 2$ the variation of the NLO cross section amounts
to only up to 2\% in all cases for both choices of $\mu_0$. The scale dependence of
the LO cross section over the entire range of $\xi$ is somewhat less pronounced for 
$\mu_0=Q$ than for $\mu_0=m_V$.
The $K$ factors, $K=\sigma_{\rm NLO}/\sigma_{\rm LO}$, are given by
$K=0.97$ for $\mu_F=\mu_R=m_V$ and  $K=1.04$ for $\mu_F=\mu_R=Q$. 
These findings indicate that both choices, $\mu_F=m_V$ and $\mu_F=Q$, in 
\sigmacLO\  yield a good estimate for the total cross section within typical
VBF 
cuts. We will see below, however, that the NLO shape of some kinematical 
distributions can be approximated better by setting $\mu_F=Q$ in the
corresponding LO predictions. 

\begin{table}[!h]
\begin{center}
\begin{tabular}{|c|c|c|c|c|}
\hline
         &\sigmacLO$(\mu_0=m_V)$&\sigmacNLO$(\mu_0=m_V)$
	 &\sigmacLO$(\mu_0=Q)$&\sigmacNLO$(\mu_0=Q)$	
	 \\
	 \hline
$\wpzjj$ &$0.224$~fb&$0.217$~fb&$0.208$~fb&$0.217$~fb 
	\\
	\hline
$\wmzjj$ &$0.122$~fb&$0.120$~fb&$0.114$~fb&$0.120$~fb
	\\
	\hline
\end{tabular}
\caption{
\label{tab:xsec}
Total cross sections for $pp\to\wpzlldec jj$ and $pp\to\wmzlldec jj$ via VBF at 
LO and NLO within the cuts of Eqs.~(\ref{eq:cutspj})--(\ref{eq:cutsmjj}) for the scale
choices $\mu_F=\mu_R=m_V$ and $\mu_F=\mu_R=Q$. The statistical errors of the quoted results are at
the permille level and therefore not given explicitly.  
}	
\vspace*{-.5cm}
\end{center}
\end{table}
For reference, we list the values of \sigmac\ at LO and NLO for $\wpzlldec jj$ 
and $\wmzlldec jj$  production in VBF for the two scale choices, $\mu_0=m_V$ and
$\mu_0=Q$ with $\xi_R=\xi_F=1$, in Tab.~\ref{tab:xsec}.
In the following, we focus on $\wpzjj$ production in VBF. The qualitative features
of EW $\wmzjj$ production are completely analogous.

\begin{figure}[!tb] 
\centerline{ 
\epsfig{figure=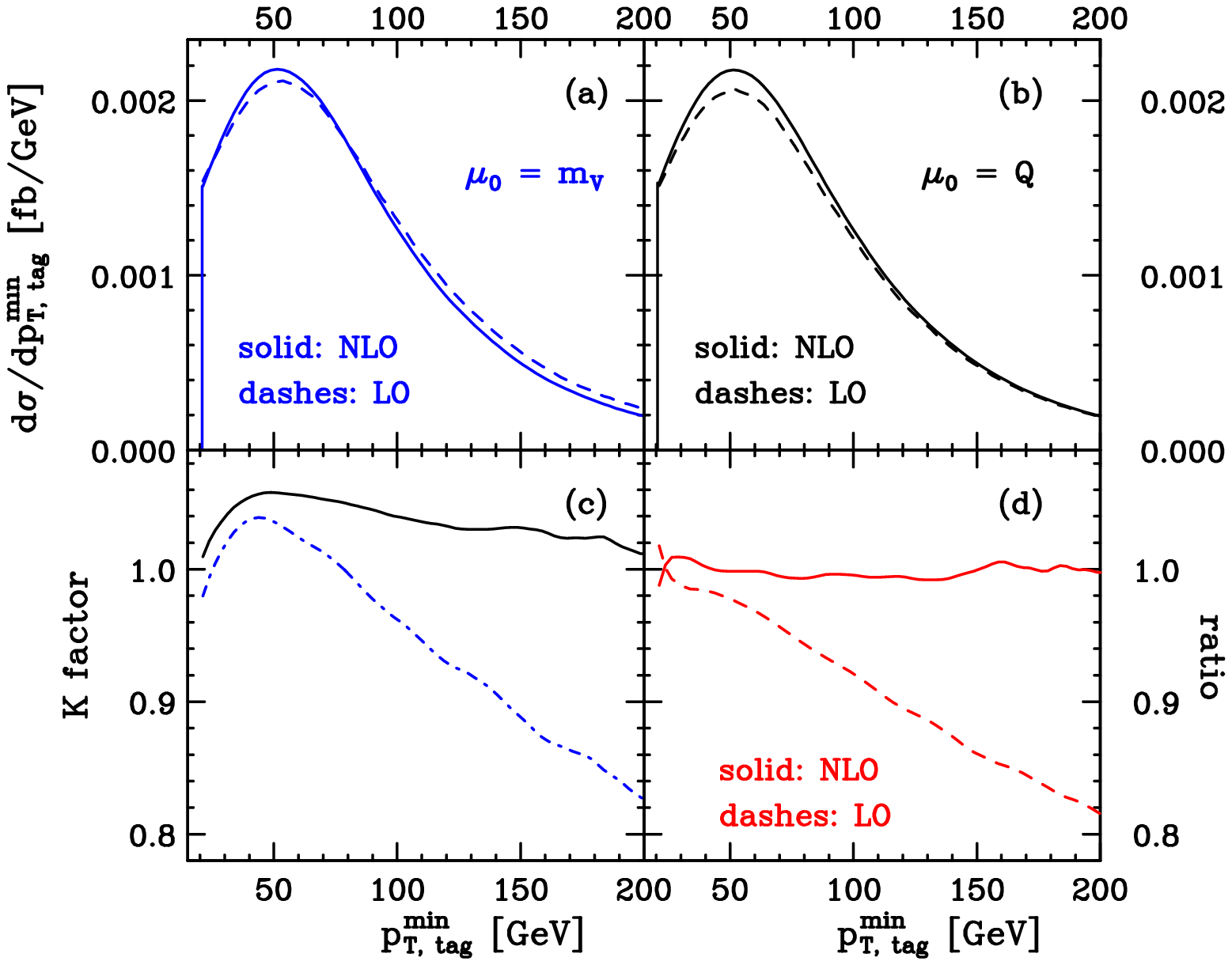,width=0.95\textwidth,clip=}
} 
\caption
{\label{fig:ptjmin}
Transverse-momentum distribution of the tagging jet with
the lowest $p_T$ in EW~$\wpzlldec jj$ production at the LHC for two different
choices of $\mu_0$ [panels~(a) and~(b)]. Panel~(c) shows the dynamical $K$
factors defined in Eq.~(\ref{eq:kfac}) for $\mu_0=m_V$ (dot-dashed blue line)
and $\mu_0=Q$ (solid black line). In  panel~(d) we have plotted the ratio
$[d\sigma/dp_{T,{\rm tag}}^{\rm min}(\mu_0=m_V)/[d\sigma/dp_{T,{\rm
tag}}^{\rm min}(\mu_0=Q)]$ at LO (dashed red line) and NLO (solid red line).
}
\end{figure}

Of particular importance for future LHC analyses are precise
predictions for jet distributions and estimates of their residual theoretical 
uncertainties.
In Fig.~\ref{fig:ptjmin} 
we therefore present the transverse momentum distribution of the tagging jet
with the lowest $p_T$ in EW~$\wpzlldec\, jj$ production for $\mu_0 = m_V$ (a) and
$\mu_0=Q$ (b) at LO and NLO (dashed and solid lines). 
Due to the cut (\ref{eq:cutsmjj}) imposed on $M_{jj}$, contributions from events with
low-$p_T$ tagging jets are suppressed and the peak of $d\sigma/dp_{T,{\rm
tag}}^{\rm min}$ is
located at about $p_T=50$~GeV. Therefore, increasing the $p_{Tj}$ cut of 
Eq.~(\ref{eq:cutspj}) to 30 or even 40~GeV would lead to only a modest 
decrease in the total cross section, an option which may be needed for high luminosity 
running.
To emphasize the impact of NLO corrections, in panel~(c) we show the 
dynamical $K$~factors, which are defined as
\beq
\label{eq:kfac}
K(x) = \frac{d\sigma_{\rm NLO}/dx}{d\sigma_{\rm LO}/dx}\;
\eeq
for the two scale choices discussed above. 
Panel~(d) illustrates the LO and NLO ratios of 
the $p_T$ distributions for $\mu_0=m_V$ and $\mu_0=Q$,
\beq
\label{eq:rat}
\frac{d\sigma/dp_{T,{\rm tag}}^{\rm min}(\mu_0=m_V)}
     {d\sigma/dp_{T,{\rm tag}}^{\rm min}(\mu_0=Q)}\;.
\eeq     
For a fixed scale $\mu_0=m_V$, the shape of the LO curve differs considerably from 
the NLO distribution, as indicated by a $K$~factor varying between $1.04$ 
and $0.8$. 
On the other hand, only a moderate change in
shape results for a dynamical scale such as $\mu_0=Q$.  The ratio of the
$p_T$ distributions with the different scales reveals that the NLO result is
barely sensitive to the choice of scale. However, setting $\mu_F=Q$ in the LO 
distributions appears to give a better estimate of the respective NLO results 
than the choice $\mu_F=m_V$. 
This feature is also present in NLO QCD calculations for deep inelastic
scattering, which constitutes the dominant part of %the QCD corrections in
the VBF configurations.  

\begin{figure}[!tb] 
\centerline{ 
\epsfig{figure=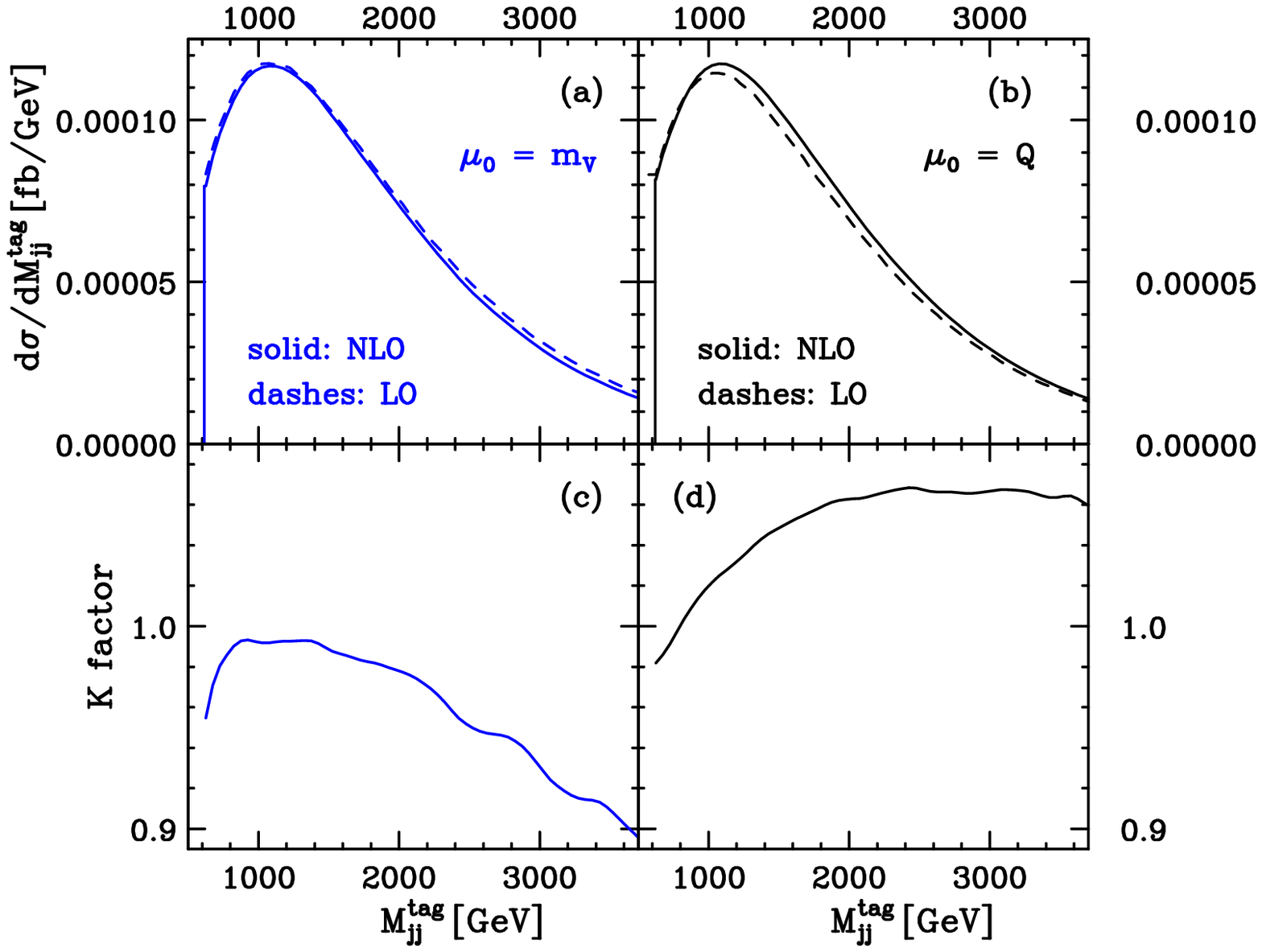,width=0.95\textwidth,clip=}
} 
\caption
{\label{fig:mjj} 
Invariant-mass distribution of the tagging jets in 
EW~$\wpzlldec jj$ production at the LHC for two different choices of $\mu_0$ 
[panels~(a) and~(b)]. Panels~(c) and~(d) show the dynamical $K$ factors
defined in Eq.~(\ref{eq:kfac}) for
$\mu_0=m_V$ (blue line) and $\mu_0=Q$ (black line).
}
\end{figure}

The invariant-mass distribution of the tagging jet pair for $\wpzlldec jj$
production is shown in Fig.~\ref{fig:mjj} 
for $\mu_0=m_V$ and $\mu_0=Q$. Due to the $M_{jj}$ cut of Eq.~(\ref{eq:cutsmjj})
the cross section vanishes for $M_{jj}<600$~GeV. 
As for the $p_T$ distribution, very similar NLO results are obtained for 
the two scale choices, while the shapes of the LO curves differ somewhat. 
For the dynamical scale choice, $\mu_0=Q$, the LO result lies below the NLO curve,
while in the case of a fixed scale, $\mu_0=m_V$, the LO curve exceeds the NLO
prediction over the entire range of $M_{jj}$. 
The different effect of the two scales can be explained by the spectrum of 
$Q$ which ranges to values higher than $m_V$. Since the parton distribution 
functions $f(x,\mu_F)$ in VBF-type reactions are probed at rather large $x$, 
in the process under investigation they decrease with increasing $\mu_F$ and are
therefore smaller for typical values of $Q$ than for $m_V$. At NLO, this
decrease is counterbalanced by respective contributions in the partonic cross
sections and the theoretical uncertainty associated with the choice of
factorization scale is reduced.

\begin{figure}[!tb] 
\centerline{ 
\epsfig{figure=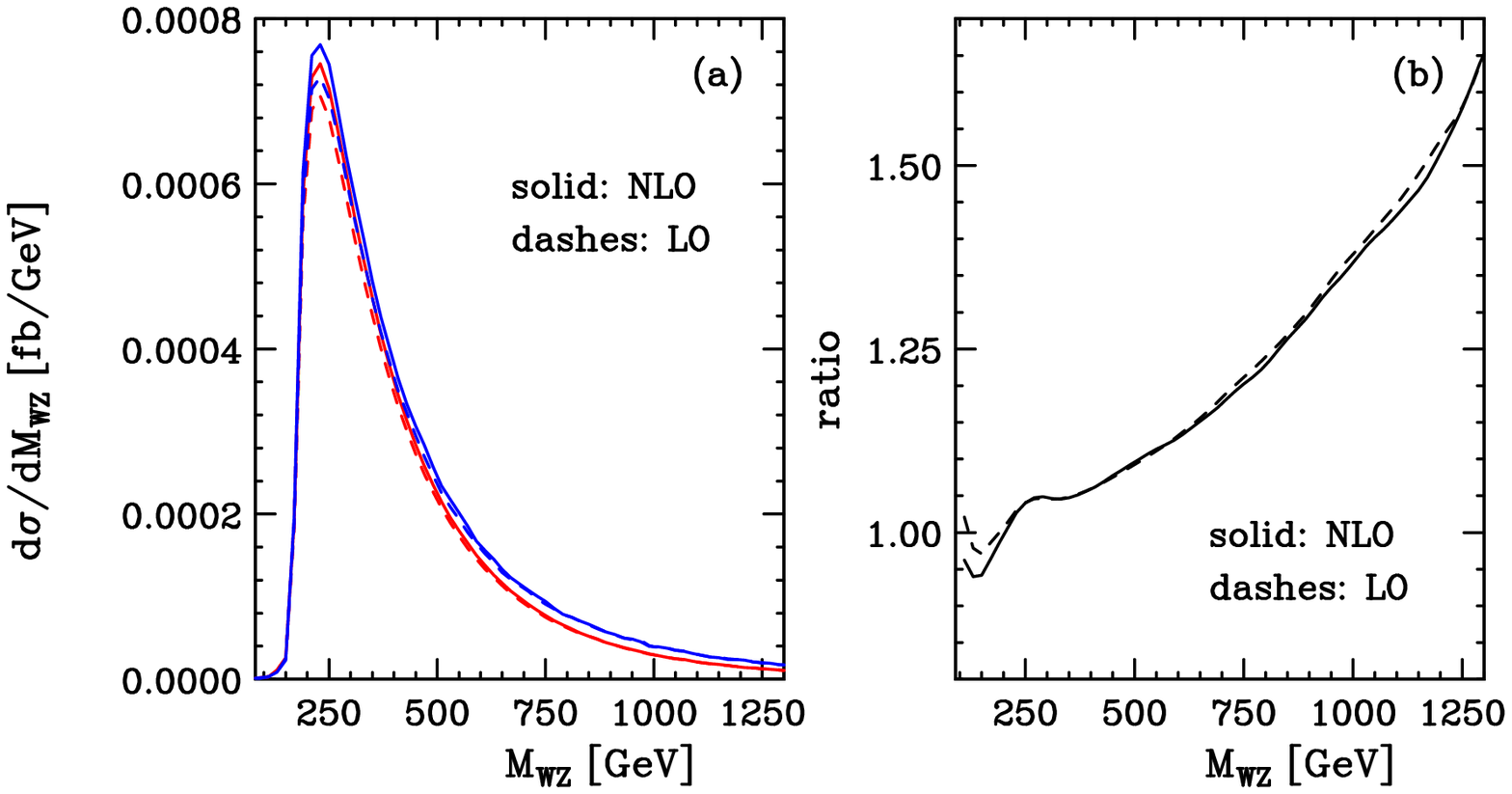,width=0.95\textwidth,clip=}
} 
\caption
{\label{fig:mwz} 
Panel~(a) shows the 
LO (dashed) and NLO (solid) distributions of the four-lepton 
invariant-mass in EW~$\wpzlldec jj$ production 
at the LHC for Higgs masses of $m_H=120$~GeV (red lower lines) and
$m_H=10$~TeV (blue 
upper lines). In panel~(b), the ratio of the distributions for the different
Higgs masses is displayed for the LO and the NLO  results. The 
factorization and renormalization scales are taken as $\mu_F=\mu_R=Q$.
}
\end{figure}

Figure~\ref{fig:mwz} 
shows the four lepton invariant mass distribution in EW~$\wpzlldec jj$  
production with $M_{WZ}$ being defined as
\beq
\label{eq:mwz}
M_{WZ} = \sqrt{(p_{e^+}+p_{\nu_e}+p_{\mu^+}+p_{\mu^-})^2} 
\eeq 
for two different values of the Higgs boson mass, $m_H=120$~GeV and $m_H=10$~TeV. 
For the latter choice, contributions from Higgs boson exchange diagrams are
effectively eliminated, i.e.\ results correspond to a ``Higgsless model''. 
Since in $\wzjj$ production processes the Higgs boson 
does not appear in the $s$-channel but only as a $t$-channel exchange boson, no
pronounced resonance behavior as, for example, in the $\zzjj$ channel, is
expected. Panel~(a)  shows that, indeed, similar shapes are
obtained for both choices of $m_H$. The ratio of the two distributions, 
\beq
 \label{eq:ratm}
\frac{d\sigma/dM_{WZ}(m_H=10~\mr{TeV})}
     {d\sigma/dM_{WZ}(m_H=120~\mr{GeV})}\;,
\eeq     
displayed in panel~(b),
reveals, however, a rise in the cross section for large $M_{WZ}$, if the Higgs
boson becomes very heavy.

%%%%%%%%%%%%%%%%%%%%%%%%%%%%%%%%%%%%%%%%%%%%%%%%%%%%%%%%%%%%%%%%%
%                                                       
\section{Conclusions}
\label{sec:concl}
In this article we have described the calculation of the NLO QCD corrections
to EW $ \wpzlldec jj$ and $\wmzlldec jj$ production at the LHC, and their
implementation in a fully flexible parton-level Monte Carlo program that
allows for the computation of kinematic distributions within realistic
experimental cuts. We found that the scale uncertainties of the NLO
predictions for total cross sections are at the 2\% level, which indicates
that the perturbative calculation is under excellent control. Care has to be
taken, however, if NLO distributions are to be approximated by LO results, as
shapes can change considerably when going from LO to NLO. In this regard, a
``proper'' choice of factorization scale can help. We found that, with a
factorization scale chosen as the momentum transfer, $Q$, of the 
$t$-channel electroweak boson, the LO calculation can better 
reproduce the shape of NLO distributions than when taking a fixed 
scale, $\mu_F=m_V$.

%
%%%%%%%%%%%%%%%%%%%%%%%%%%%%%%%%%%%%%%%%%%%%%%%%%%%%%%%%%%%%%%%%%
%
\section*{Acknowledgments}
We are grateful to the {\tt MadGraph/MadEvent} group, who gave us 
access to their computer cluster for performing the comparison with 
{\tt MadEvent} generated code.
B.~J.\ wishes to thank the RIKEN research center at Wako for kind hospitality and
support.
This research was supported in part by the Deutsche Forschungsgemeinschaft
under SFB TR-9 ``Computergest\"utzte Theoretische
Teilchenphysik'' and in part by INFN.
%
%%%%%%%%%%%%%%%%%%%%%%%%%%%%%%%%%%%%%%%%%%%%%%%%%%%%%%%%%%%%%%%%%
%

\end{document}